\documentclass{WileyMSP-template}

\usepackage{PRIMEarxiv}

\usepackage{amsmath}
\usepackage{xcolor}
\usepackage[utf8]{inputenc} 
\usepackage[T1]{fontenc}    
\usepackage{hyperref}       
\usepackage{url}            
\usepackage{booktabs}       
\usepackage{amsfonts}       
\usepackage{nicefrac}       
\usepackage{microtype}      
\usepackage{lipsum}
\usepackage{fancyhdr}       
\usepackage{graphicx}       
\graphicspath{{media/}}     
\usepackage{tikz}
\usetikzlibrary{arrows.meta,
                chains,
                positioning,
                shapes.geometric
                }

\hypersetup{
    colorlinks,
    linkcolor={red!50!black},
    citecolor={blue!50!black},
    urlcolor={blue!80!black}
} 
\begin{document}
\title{Machine Learning for maximizing the memristivity \\ of single and coupled quantum memristors}

\author{Carlos Hernani-Morales$^{1,2}$, Gabriel Alvarado$^3$, Francisco Albarr\'an-Arriagada$^{4,5}$, \\ Yolanda Vives-Gilabert$^2$,
Enrique Solano$^3$, Jos\'e D. Mart\'in-Guerrero$^{2,6}$\\$^{1}$ Quantum Spain, Universitat de València, \\C/Dr. Moliner, 50, 46100 Burjassot, Valencia. Spain. \\$^{2}$IDAL, Electronic Engineering Department,  ETSE-UV, Universitat de València, \\Avgda. Universitat s/n, 46100 Burjassot, Valencia, Spain\\
$^{3}$Kipu Quantum, Greifswalderstrasse 226, 10405 Berlin, Germany \\$^{4}$Departamento de F\'isica, Universidad de Santiago de Chile (USACH), Avenida V\'ictor Jara 3493, 9170124, Santiago, Chile. \\$^{5}$Center for the Development of Nanoscience and Nanotechnology, 9170124, Estaci\'on Central, Chile. \\$^{6}$Valencian Graduate School and Research Network of Artificial Intelligence (ValgrAI), Spain\leavevmode\\carlos.hernani@uv.es, jose.d.martin@uv.es}

\newpage

\maketitle

\begin{abstract}

We propose machine learning (ML) methods to characterize the memristive properties of single and coupled quantum memristors. We show that maximizing the memristivity leads to large values in the degree of entanglement of two quantum memristors, unveiling the close relationship between quantum correlations and memory. Our results strengthen the possibility of using quantum memristors as key components of neuromorphic quantum computing.

\end{abstract}


\section{Introduction} \label{section:introduction}

The memristor was proposed as the fourth basic circuit element, characterized by a flux-charge relationship \cite{1083337,1454361}. This device exhibits rich nonlinear properties and it is distinguished by a pinched hysteresis curve in the current-voltage (I/V) plane, which can be described by Kubo’s response theory~\cite{doi:10.1143/JPSJ.12.570}. Since the experimental implementation of a memristor in a doped semiconductor by HP Labs in 2008~\cite{Strukov2008}, memristors have garnered significant interest in several areas, including analog computing~\cite{Barrios2019} and neuromorphic computing~\cite{schuman2017survey}. A notable application of memristors is the design of devices that mimic biological neural synapses \cite{Wang2017} and neural networks \cite{doi:10.1080/00018732.2010.544961}.
Furthermore, memristor-enabled neuromorphic computing goes beyond the traditional von Neumann computing paradigm, avoiding the von Neumann bottleneck, which is one of the fundamental limitations of current classical computers \cite{Li_2018,Markovic2020,Islam_2019}.

Quantum computing \cite{GYONGYOSI201951} aims to revolutionize computation by exploiting exclusively quantum phenomena to surpass the capabilities of classical computers, as we can see from recent breakthroughs~\cite{Arute2019,doi:10.1126/science.abe8770, PhysRevLett.127.180501,doi:10.1126/sciadv.abi7894,morvan2023phase}. The merge between quantum computing and neuromorphic computing or neuromorphic quantum computation (NQC) \cite{PhysRevA.104.062605,PhysRevApplied.18.034004,10.1063/5.0020014} seeks to implement brain-inspired devices with quantum hardware and software, and may lead to groundbreaking technologies. Therefore, it is natural to wonder whether a quantum version of memristors, i.e., quantum memristors (QMs)~\cite{Pfeiffer2016-oj}, could become a significant component of neuromorphic quantum hardware similar to its classical counterpart. This question has motivated the theoretical proposal of quantum memristors in different platforms, such as quantum photonics \cite{10.1063/1.5036596,ma13040864}, superconducting circuits \cite{Pfeiffer2016-oj,PhysRevApplied.6.014006, PhysRevApplied.10.014013, PhysRevApplied.2.034011,Salmilehto2017-zi}, ion traps \cite{e25081134}, and light-matter \cite{Tang2023,PhysRevApplied.17.024056}. Some recent experimental realizations are especially remarkable \cite{Spagnolo2022-ie,GAO2022100007}.

The use of quantum memristors for NQC applications need a deep understanding of how they can be connected and correlated. However, most proposals have been limited to single quantum memristor models, and further study is required for the case of coupling and scaling quantum memristors. In this context, \cite{PhysRevA.104.062605} studied memristive dynamics of two-coupled quantum memristors concerning their entanglement properties and \cite{PhysRevApplied.18.034004} a tripartite entanglement. These works show that the memristive properties and entanglement have an opposite behavior, providing the signs of a nontrivial role of quantum correlations in memristive devices. Nonetheless, there is more to explore in this direction since increasing the number of subsystems introduces quantum correlations that cannot be observed in bipartite systems, such as multipartite entanglement and entanglement monogamy \cite{PhysRevA.61.052306, PhysRevLett.113.100503}. Addressing this point can be relevant for future applications, such as reservoir quantum computing (RQC) \cite{PhysRevApplied.8.024030} where arrays of quantum memristors could compose the reservoir \cite{Spagnolo2022-ie} which will require scaling quantum memristor arrays into quantum neural networks and neuromorphic quantum computers, as well as the design of devices with optimal memristive properties.

The optimal configuration of single and coupled QMs is needed for neuromorphic quantum technologies. During the last years, ML has been increasingly used for almost every conceivable problem where an optimal response is necessary, e.g.,  medical image classification \cite{FERRERSANCHEZ2022105967}, chat-bots like ChatGPT \cite{liu2023summary}, image generation by stable diffusion \cite{rombach2022highresolution}, finance \cite{PhysRevResearch.3.013167, e25020323}, and more.  In physics,  some examples of application of ML include state characterization~\cite{Wang2021-vr},  noise reduction in CERN experiments \cite{Flores} and faster simulations of PDE systems using Physics Informed Neural Networks (PINNs)~\cite{ferrersanchez2023gradientannihilated}. Due to the flexibility of ML techniques to solve a wide range of problems, seems to be natural the use of ML to design efficient memristive devices, approaching the usefulness of quantum memristors.

In this work, we study the use of ML models to find the optimal set of parameters which maximizes the memristive behavior of  a single QM and a system composed of two-coupled QMs. We found that in the optimal case, there is strong evidence that supports the connection between quantum correlations and the memristive behavior. Specifically, an enhancement in the memristive properties, also means an enhancement in the quantum correlation. This helps to characterize quantum memristive devices accurately which will, in turn, take one step closer to implementable and useful neuromorphic quantum computing.

The rest of the paper is organized as follows.  Sec. \ref{section:quantum_memristor} deals with the models of single quantum memristor and two-coupled memristors.  In Sec. \ref{section:methodology}, we describe the ML pipeline used for both cases, the ensemble methods used (Random Forests, Extra Trees, XGBoost and LightGBM),  and finally,  the data set generation.
Sec. \ref{section:results} reports the performance of the proposed models in terms of regression metrics, ending up the paper with the main conclusions of our work and proposals for further research.

All the code used, both for data generation and machine learning, is available at \url{https://github.com/carlos-hernani/QuantumMemristor}.


\section{Quantum Memristors}
\label{section:quantum_memristor}
The memristor is a two-terminal device with resistive memory effects, in which the resistance depends on the history of charges across the device. The ideal memristor initially proposed by L. Chua is completely described by its state-dependent Ohm's law \cite{1083337}. Later the concept of memristor was extended to a more general class of systems called memristive systems \cite{1454361}, defined by the following equations:
\begin{subequations}
\begin{align}
y &= f(s,u,t)u \, , \\
\dot{s} &= g(s,u,t) \,,
\end{align}
\end{subequations}
where $u$ and $y$ denote input and output variables, respectively, and $s$ denotes a state variable. All variables are assumed to be dependent on time. The equations that characterize the ideal memristor are recovered by setting $f(s,u,t)~=~f(s)$ and $g(s,u,t)=g(u)$. The memristive equations written in this form are also called a state dependent Ohm’s law when $u$ and $y$  correspond to current and voltage, and $f$ is called memductance (memristance) when the input variable has units of current (voltage).

Quantum memristors are obtained by implementing the characteristic equations of a classical memristor with quantum observables of a quantum system. These equations can be the state dependent Ohm’s law, or the generalized version for memristive systems. In principle, any device that satisfies these equations can be considered a quantum memristor, therefore they are not bound to any specific platform. We can define a quantum memristor analogously to their classical counterpart. A quantum memristor is a quantum system with observables $\hat{y}$  and $\hat{u}$ whose expectation values $\langle \hat{y} \rangle$ and $\langle \hat{u} \rangle$ follow the relationship
\begin{subequations}
\begin{align}
\langle \hat{y} \rangle &= f(\langle \hat{x} \rangle, \langle \hat{u} \rangle,t) \langle \hat{u} \rangle \, , \\
\langle\dot{ \hat{x}}\rangle &= F\big(\langle \hat{x} \rangle, \langle \hat{u} \rangle, t \big) \, .
\end{align}
\end{subequations}
Here, $  G\big(\langle \hat{x} \rangle, \langle \hat{u} \rangle, t \big)$ and $ F\big(\langle \hat{x} \rangle, \langle \hat{u} \rangle, t \big)$ are the quantum analog of the response and state variable function, respectively.

\begin{figure}
    \centering
    \includegraphics[width=0.7\linewidth]{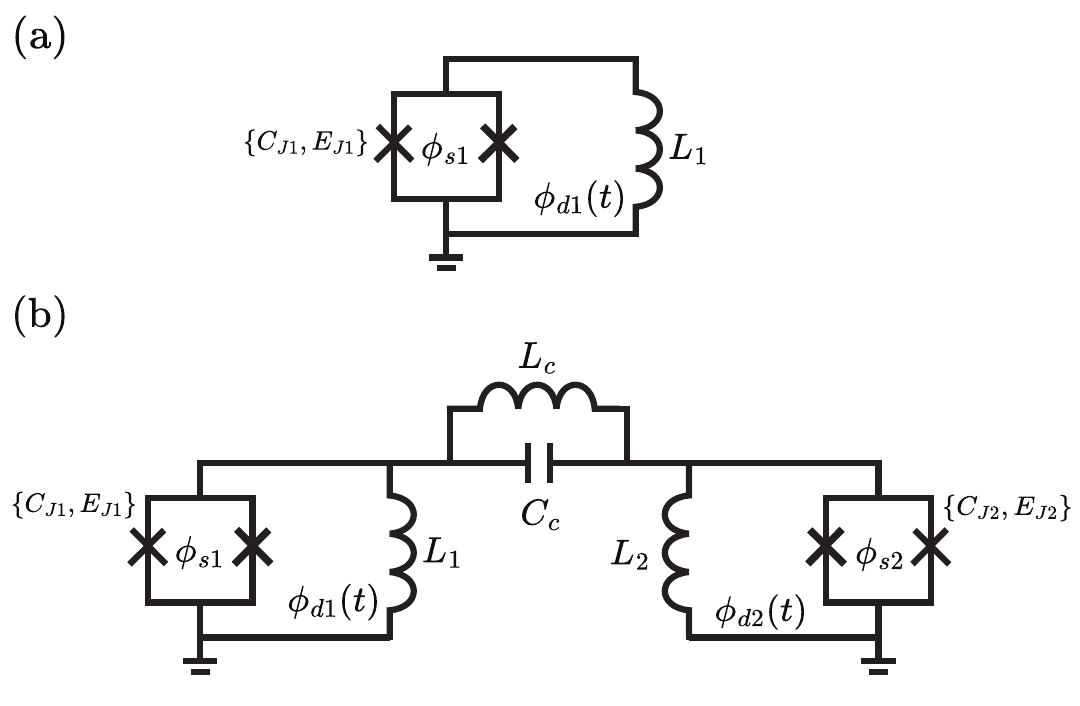}
    \caption{Circuit diagram of (a) a single quantum memristor and (b) two quantum memristors coupled by either a capacitor, with capacitance $C_c$, or by an inductor, of inductance $L_c$.}
    \label{fig.1}
\end{figure}

We consider quantum memristors composed of a conductance-asymmetric superconducting quantum interference device (CA-SQUID) connected in parallel to an inductor $L_{\ell}$. This forms a closed loop that is threaded by an external signal $\phi_{d\ell}$, as shown in Fig.~\ref{fig.1}(a). The Josephson junctions in the SQUID have two current contributions, a nondissipative current due to tunneling of Cooper pairs \cite{JOSEPHSON1962251} and a dissipative current due to the tunneling of quasiparticles \cite{PhysRevB.86.184514}, which is of memristive nature. The Josephson junctions in the SQUID are made with different materials such that they have different conductance but the same critical current \cite{PhysRevApplied.2.034011}. In this way, by applying a bias flux of half a flux quantum $\Phi_0/2$ through the SQUID loop we can completely suppress the effective critical current. Additionally, we apply a bias flux $\phi_{d1}(t)$ on the loop enclosed by the inductance $L_1$. Furthermore, two quantum memristors as described above can be coupled by either a capacitor or an inductor as shown in Fig.~\ref{fig.1}(b), such as system is described by the following Hamiltonian (for a detailed derivation see refs.~\cite{PhysRevA.104.062605}):
\begin{equation}
H = \sum_{\ell=1,2} \big[ E_{C, \ell} \hat{n}_{\ell}^2 + \frac{E_{L, \ell}}{2}\hat{\phi}_{\ell}^2 \big] + E_{C, 1, 2}  \hat{n}_{1} \hat{n}_{2} - E_{L, 1, 2} \hat{\phi}_{1} \hat{\phi}_{2},
\end{equation}
where $\hat{n}_{\ell}$ and $\hat{\phi}_{\ell}$ represent the dimensionless charge and phase operators of the $\ell$th quantum memristor. ${E_{C, \ell} = 2e^2C^{-1}_{\ell, \ell}}$ and $E_{C, 1, 2} = 2e^2C^{-1}_{1, 2}$ are the charge energy with $C^{-1}_{j,k}$ the $(j,k)$ element of the inverse of the capacitance matrix. $E_{L, \ell} = 2\varphi_0L^{-1}_{\ell, \ell}$ and $E_{L, 1, 2} = 2\varphi_0L^{-1}_{1, 2}$ are the inductive energy with $L^{-1}_{j,k}$ the $(j,k)$ element of the inverse of the inductance matrix, and $\varphi_0 = \hbar/2e$ is the reduced quantum flux. 

Redefining the operators in terms of creation and annihilation operators as
\begin{eqnarray}
\hat{n}_{\ell} = \frac{i}{4g_{\ell}} \big( \hat{a}_{\ell}^{\dagger}  - \hat{a}_{\ell}\big) \nonumber \, , \\
\hat{\phi}_{\ell} = 2g_{\ell} \big( \hat{a}_{\ell}^{\dagger}  + \hat{a}_{\ell}\big),  
\end{eqnarray}
where $g_{\ell} = \big(E_{C,\ell}/32 E_{L,\ell} \big)$, the Hamiltonian take the form
\begin{equation}
\label{Hamiltonian2q}
\hat{H} = \sum_{\ell=1,2} \hbar \omega_{\ell} \hat{a}_{\ell}^{\dagger} \hat{a}_{\ell} - \sqrt{\omega_1 \omega_2}(\alpha - \beta) \big(\hat{a}_1^{\dagger} \hat{a}_2  + \hat{a}_2^{\dagger} \hat{a}_1\big) \, ,
\end{equation}
with $\omega_{\ell} = \sqrt{2E_{C,\ell} E_{L,\ell}}/\hbar$ the frequency of the $\ell$th QM.  Here, $\alpha = E_{L,1}/ \sqrt{E_{L,1} E_{L,2}}$ and ${\beta= E_{C,1}/ \sqrt{E_{C,1} E_{C,2}}}$. The energy loss of the admittance is described as a quasiparticle bath at zero temperature; this mechanism can be modeled with the time-dependent master equation ($\hbar=1$):
\begin{equation}
\label{mastereq}
\dot{\rho}(t) = i\big[ \hat{H}, \rho \big] + \sum_{\ell=1,2} \frac{\Gamma_{\ell}(t)}{2} \bigg(  \hat{a}_{\ell}\rho \hat{a}_{\ell}^{\dagger} - \frac{1}{2} (\hat{a}_{\ell}^{\dagger} \hat{a}_{\ell} \rho + \rho \hat{a}_{\ell}^{\dagger} \hat{a}_{\ell} )  \bigg) \, ,
\end{equation}
where $\hat{H}$ is the Hamiltonian of Eq.~(\ref{Hamiltonian2q}) and $\Gamma_{\ell}(t) = g_{\ell}^2 \omega_{\ell} \exp(-g_{\ell}^2)\big((1 + \cos[\phi_{d\ell}(t)])/2 \big)$ is the time-dependent decay rate, where we have considered spectral density of the quasiparticle bath $S_{QP}(\omega_{\ell}) = \lambda \omega_{\ell}$ and the fluxoid quantization condition in the outer loop. We note that this time-dependent decay rate can be driven by the external magnetic flux $\phi_{d\ell}(t) = \phi_{0, \ell} + A\sin(\omega_{\ell} t)$. For more details about the effective decay rate see Ref.~\cite{Salmilehto2017-zi}.

From the master equation above we can obtain the equations of motion for the number and phase operator as follows
\begin{eqnarray}
\frac{d}{dt}\langle \hat{n}_{\ell} \rangle = E_{L, \ell} \langle \hat{\phi}_{\ell} \rangle -  E_{L, 1,2} (\delta_{1, \ell}  \langle \hat{\phi}_{2} \rangle - \delta_{2,\ell}  \langle \hat{\phi}_{1} \rangle) - \frac{\Gamma_{\ell}}{2}  \langle \hat{n}_{\ell} \rangle \, , \\
\frac{d}{dt}\langle \hat{\phi}_{\ell} \rangle = E_{C, \ell} \langle \hat{n}_{\ell} \rangle -  E_{C, 1,2} (\delta_{1, \ell}  \langle \hat{n}_{2} \rangle - \delta_{2,\ell}  \langle \hat{n}_{1} \rangle) - \frac{\Gamma_{\ell}}{2}  \langle \hat{\phi}_{\ell} \rangle \, .
\end{eqnarray}

From the equations of motion we obtain the memristive variables, corresponding to the quasiparticle current  $\hat{I}_{\textrm{QP}, \ell}(t)$ and voltage across the capacitor $\hat{V}_{\textrm{cap}, \ell}(t)$ which can be expressed as:
\begin{eqnarray}
    \label{memcurrent}
    \hat{I}_{\textrm{QP}, \ell}(t) = G_{\ell}(t) \hat{V}_{\textrm{cap}, \ell} (t) ,\\
    \label{memvoltage}
    \hat{V}_{\textrm{cap}, \ell}(t) =-2e\langle \hat{n}_{\ell}(t) \rangle/C_{\Sigma, \ell},
\end{eqnarray}
where $G_{\ell} = C_{\Sigma, \ell} \Gamma_{\ell}(t)/2$ is the memductance of each quantum memristor. The dynamics of the memristive variables define an hysteretic response characteristic of memristive behavior which depends on the initial state. In addition, it is assumed that the quantum memristors operate in the two-level approximation, ensured by the choice of initial state, which is parameterized as $\vert \Psi_{\ell} (\theta_{\ell}, \eta_{\ell})\rangle = \cos(\theta_{\ell}/2) \vert 0 \rangle + e^{i \eta_{\ell}}\sin(\theta_{\ell}/2)\vert 1 \rangle$.

The system evolves according to Eq.~(\ref{mastereq}) for a given initial state, and the memristive variables define a pinched  hysteresis loop in the plane I/V. The area enclosed by the hysteresis loop is related to the memory effects of the device \cite{IEEE2012Biolek, EL2014Biolek}, which can be characterized by the form factor, defined as the ratio of the area enclosed by the hysteresis loop, $A$, and its corresponding perimeter $P$:
\begin{equation}
\label{form_factor_eqn}
\mathcal{F} = 4\pi\frac{A}{P^2} 
\end{equation}
which is invariant under scaling allowing to compare the memory effects in the I/V of any memristor.

In the case of a single quantum memristor, the memristive behavior and its corresponding hysteresis loop are mainly determined by two factors: the choice of the initial state v$\vert \Psi (\theta, \eta)\rangle$, and the amplitude of the spectral density $S_{QP}(\omega)$ \cite{PhysRevA.104.062605}. When considering coupled quantum memristors, the memristive behavior is also influenced by the coupling capacitance $C_c$ and inductance $L_c$. In particular, for coupled quantum memristors, the memristive behavior of each subsystem is highly sensitive to both the circuit parameters and the initial state. Furthermore, it has been shown that for particular instances, the quantum correlations between the coupled quantum memristors is directly related to the form factor, and therefore, the memristive behavior of the system. To the best of our knowledge, there has not yet been a study on the optimal configuration that maximizes the memristive behavior and how it relates to quantum correlations.

\section{Proposed methodology}\label{section:methodology}

The main objective of this work is to analyze the dependence of the memristive behavior, characterized by the form factor, on the circuit parameters and to find its optimal configuration that maximizes the form factor. In the case of coupled quantum memristors, we are also interested in understanding the relationship between the optimal and sub-optimal configurations and the quantum correlations generated between the memristors. 

For a single memristor, we consider the amplitude of the spectral density, represented by $\lambda$, and the angles $\theta$ and $\phi$ that determine the initial state according to the Bloch sphere parametrization. In the case of coupled quantum memristors, we consider these quantities for each quantum memristor, represented by $\lambda_{\ell}$, $\theta_{\ell}$, $\phi_{\ell}$ as well as the capacitive and inductive coupling $C_c$ and $L_c$. From previous results, it has been determined that setting $\theta = \pi/2$ yields the highest values of the form factor \cite{PhysRevA.104.062605}. As a result, we fix $\theta$ to this value throughout the rest of this work.

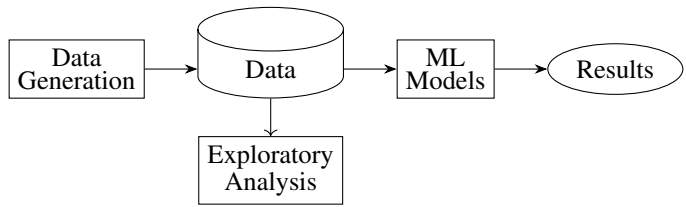
\begin{figure}[h]
\begin{center}
\begin{tikzpicture}[
    node distance = 5mm and 7mm,
      start chain = going right,
 disc/.style = {shape=cylinder, draw, shape aspect=0.3,
                shape border rotate=90,
                text width=17mm, align=center, font=\linespread{0.8}\selectfont},
  mdl/.style = {shape=ellipse, aspect=2.2, draw},
  alg/.style = {draw, align=center, font=\linespread{0.8}\selectfont}
                    ]

    \begin{scope}[every node/.append style={on chain, join=by -Stealth}]
\node (n1) [alg] {Data\\ Generation};
\node (n2) [disc] {Data};
\node (n3) [alg] {ML\\Models};
\node (n4) [mdl] {Results};
    \end{scope}
\node[below=of n2] (EA) [alg] {Exploratory\\Analysis};
\draw [->] (n2) -- (EA);
    \end{tikzpicture}
\end{center}
    \caption{The basic outline of the ML pipeline.}
    \label{fig:ml_pipeline}
\end{figure}

To tackle this problem we follow the pipeline shown in Fig. \ref{fig:ml_pipeline}. The first step is data generation (Sec.~\ref{subsection:data set_gen}). Once we have the data we can start its exploratory analysis (Secs.~\ref{subsection:single_qm} and \ref{subsection:two_coupled_qm}) before implementing the ML models (Sec.~\ref{subsection:ml_models}). The results of these models are detailed in Sec. \ref{section:results}.

As explained in Sec.~\ref{section:quantum_memristor},  quantum memristors have different tunable parameters. Our proposal is to make use of ML to find an accurate relationship between the parameters and the memristor behavior,  in order to maximize memristivity, indirectly estimated by means of the form factor.  This optimization leads, in turn,  to a strong connection between quantum correlations and the form factor.

The main characteristics of a QM are its quantumness and its hysteresis loop. The area enclosed by the hysteresis curves can be related to the memory effects of a memristor. Besides, the physical properties and the maximum values of the voltage (input variable) are related to the perimeter of the hysteresis curve. In line with this, a good candidate to evaluate the memristive properties of the memristor is the form factor given by Eq.~(\ref{form_factor_eqn}).

The form factor is a dimensionless quantity that measures the area enclosed by a given perimeter, where its maximal value $\mathcal{F}=1$ is obtained for a circle (maximal area). Meanwhile, the minimal value $\mathcal{F}=0$ is obtained for a line (minimal area). It allows us to compare the different curves without caring about the losses in the maximum voltage value in each loop.

\subsection{Data set generation} \label{subsection:data set_gen}

Using the code for data generation at \url{https://github.com/carlos-hernani/QuantumMemristor} that implements Eq.~(\ref{mastereq}) and obtains the memristive variables from Eq.~(\ref{memcurrent}) and Eq.~(\ref{memvoltage}), a random\footnote{A random seed is used for reproducible results.} grid of samples is calculated, from which the form factor will be derived. There are two data sets, one for the case of a single quantum memristor and another for the case of two-coupled quantum memristors. 

\subsection{Single Quantum Memristor} \label{subsection:single_qm}

For the case of a single quantum memristor, we set $C_c$ and $L_c$ to zero and generated a data set of 2,000 samples with two QM parameters, $\lambda$ and $\phi$, using the form factor as target variable by calculating the time evolution of the system with up to 10 periods of the driving signal $T = 2\pi/\omega$.
The statistical description of the data set is shown at Table \ref{table:single_qm_stat_descrip}.  

\begin{table}[h]
  \caption{Quartile description of single quantum memristor parameters and form factor.}\label{table:single_qm_stat_descrip}
\centering
    \begin{tabular}{cccc} \toprule
    {}  & $\phi$ & $\lambda$ & Formfactor \\
\midrule
count & 2000 & 2000 & 2000 \\
mean & 3.085177 & 50.115547 & 0.181724 \\
std & 1.787681 & 29.181347 & 0.085811 \\
min & 0.013517 & 0.176952 & 0.001295 \\
25\% & 1.496520 & 24.855546 & 0.124916 \\
50\% & 3.133857 & 49.663515 & 0.177362 \\
75\% & 4.617136 & 75.723366 & 0.252894 \\
max & 6.219332 & 99.978500 & 0.324210 \\
\bottomrule
\end{tabular}   
\end{table}

\begin{figure}[h]
    \centering
    \includegraphics[width=1.1\textwidth]{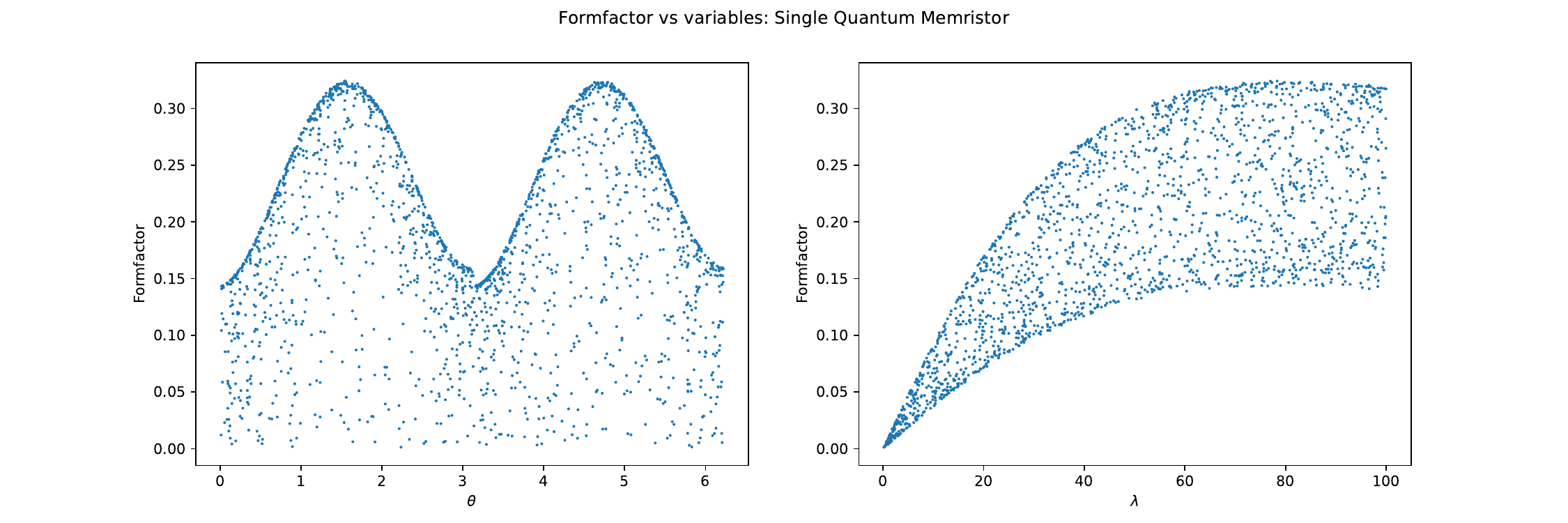}
    \caption{Form factor plotted versus the QM parameters.}
    \label{fig:singleqm_plot_ff_vs_params}
\end{figure}

We also plot the form factor versus the QM parameters to see their dependency at Fig \ref{fig:singleqm_plot_ff_vs_params}. There is a straightforward  relationship between the form factor and the parameters which, in turn, makes it an attainable problem to solve by machine learning, as we will see at Sec. \ref{section:results}.

\newpage
\subsection{Two-Coupled Memristors} \label{subsection:two_coupled_qm}

For the case of two-coupled quantum memristors, we generated a data set of 15,000 samples with four QM parameters and the target variables, which are the form factors for both QMs (in this case they are the same as they share the same parameters). We evaluate the form factor by calculating the time evolution of the system with up to 10 periods of the driving signal $T = 2\pi/\omega_{\ell}$. To make sure that the capacitance and the inductance can have zero values we used a parameter grid of 10x10x10x10.
The statistical description of the data set is shown at Table \ref{table:two_qm_stat_descrip}. 

\begin{table}[h]
  \caption{Quartile description of the two-coupled QMs parameters and form factor.}
    \label{table:two_qm_stat_descrip}
\centering
    \begin{tabular}{ccccccc} \toprule
    {} & $C_{12}$ & $L_{12}$ & $\theta$ & $\lambda$ & Form Factor 1 & Form Factor 2 \\
\midrule
count & 15000 & 15000 & 15000 & 15000 & 15000 & 15000 \\
mean & 1e-12	& 1e-08	&3.12	& 50.2	& 0.231	& 0.231 \\
std & 6.18e-13	& 6.19e-09	& 1.92	& 31	& 0.0507	& 0.0507 \\
min & 0	& 0	& 0.01	& 0.1	& 0.132	& 0.132 \\
25\% & 4.44e-13	& 4.44e-09	& 1.39	& 22.3	& 0.194	& 0.194 \\
50\% & 1e-12	& 9.98e-09	& 3.12	& 51	& 0.225	& 0.225 \\
75\% & 1.56e-12	& 1.56e-08	& 4.84	& 77.8	& 0.263	& 0.263 \\
max & 2e-12	& 2e-08	& 6.22	& 100	& 0.503	& 0.503 \\
\bottomrule
\end{tabular}
   
\end{table}

\begin{figure}[h]
    \centering
    \includegraphics[width=\textwidth]{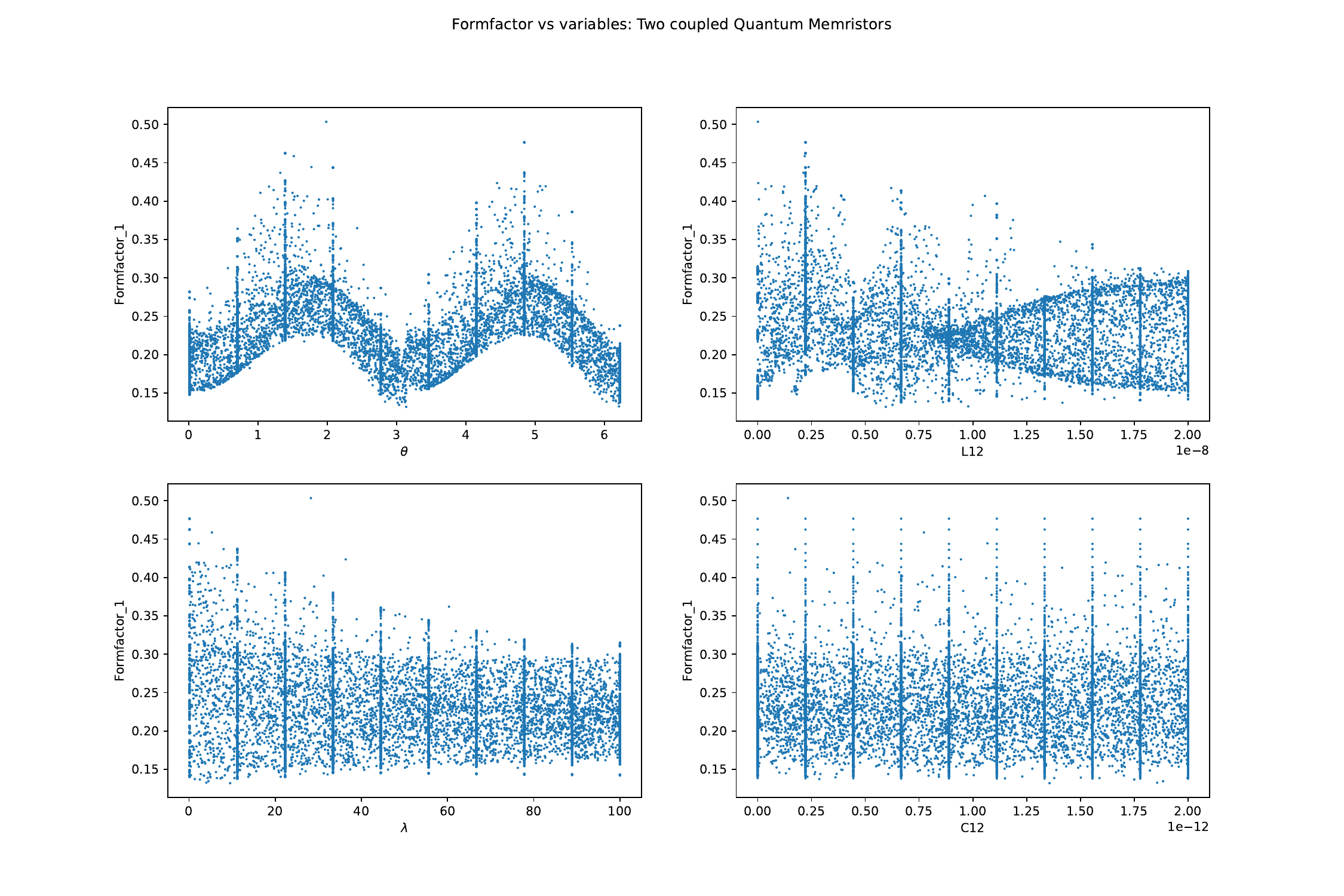}
    \caption{Form factor plotted versus all the parameters of the system of two coupled QM.}
    \label{fig:twoqm_plot_ff_vs_params}
\end{figure}

We also plotted the form factor versus all parameters to see their dependency at Fig \ref{fig:twoqm_plot_ff_vs_params}.Now we can see that the noise is more relevant than for one QM but it is still clear that there is a relationship between form factor and parameters. ML still perform well but not as accurately, as in the case of the single QM case, as shown in Sec. \ref{section:results}.

\newpage
\subsection{Machine Learning Models} \label{subsection:ml_models}

The first model used for the case of a single quantum memristor was a Support Vector Regressor (SVR)~\cite{NIPS1996_d3890178} using a Gaussian Kernel. This case is sufficiently simple that a SVR can achieve very good results in the $R^2$-score for the test data set, even though these models are especially adequate when the feature space is large compared to the number of examples \cite{NIPS1996_d3890178}. SVRs are a generalization of Support Vector Machines (SVMs)~\cite{NIPS1996_d3890178}, which try to solve classification problems by formulating them as convex optimization problems. In this manner, one has to find the suitable hyperplanes that classify correctly as many training samples as possible. In particular, SVRs work by creating a transformed data space in which the problem is more easily solvable and, ideally the problem is transformed into a linear one. That transformation between spaces is carried out by the so-called kernels. Gaussian, linear, and polynomial kernels were used in this experimentation. An SVR introduces a region in the hyperspace of the problem called $\epsilon$-tube, within which, all predictions are considered as correct.

When dealing with two-coupled quantum memristors,  more complex models were necessary to describe the behavior of the system accurately; in particular, we focused on tree-based ensemble methods.
Ensemble methods involve combining the prediction of multiple individual models, known as base models, to generate a final prediction \cite{MOHAMMED2023757, Ganaie_2022}. These methods have proven to be an effective strategy for achieving higher accuracy and robustness. Ensemble methods offer several advantages over single models, including improved generalization, reduced overfitting, and increased stability \cite{Dong2020}. By combining the predictions from multiple models, ensemble methods can effectively mitigate biases and errors present in individual models, resulting in more robust and accurate predictions.
The rest of this Section is dedicated to briefly describe four ensemble methods.

\subsubsection{Random Forest} \label{subsubsection:random_forest}

Random Forests (RFs) are a combination of tree predictors \cite{decision_trees} such that each tree depends on the values of a random vector sampled independently and with the same distribution for all trees in the forest \cite{Breiman2001}.
The underlying principle behind RFs is that, being an ensemble method, the combination of multiple decision trees, where the splits are chosen to be optimal, helps in reducing overfitting and improves the overall predictive performance. By leveraging the randomness in both data sampling and feature selection, RFs achieve a balance between variance and bias, making it robust and effective in handling complex problems.

\subsubsection{Extra-Trees} \label{subsubsection:extra_trees}

Extra-trees \cite{Geurts2006}, also called Extremely Randomized Trees (ERTs),  is another tree-based ensemble method where the algorithm builds an ensemble of trees where the node splitting is done fully at random and uses the whole learning sample, rather than a subset, to grow the trees~\cite{Geurts2006}.

The main differences between ERTs and RFs are that ERTs make use of the whole data set to train trees whereas RFs is based on bagging, i.e. , using randomly selected subsets of the whole training set with repetition. Also, ERTs make the node splits fully at random in contrast to RFs where the splits are done using an optimality criteria. This last difference makes ERTs faster to train.

\subsubsection{Extra-Gradient Boosting} \label{subsubsection:xgboost}
Extra-Gradient Boosting (XGBoost) is a popular and efficient open-source implementation of the gradient-boosted decision tree algorithm (GBDT) that combines boosting and gradient-descent techniques to produce accurate predictive models. Gradient boosting is a supervised learning algorithm, which attempts to accurately predict a target variable by combining the estimates of a set of simpler, weaker models.
When using gradient boosting for regression, the weak learners are regression trees, and each regression tree maps an input data point to one of its leaves that contains a continuous score. XGBoost minimizes a regularized (L1 and L2) objective function that combines a convex loss function (based on the difference between the predicted and target outputs) and a penalty term for model complexity (in other words, the regression tree functions). The training proceeds iteratively, adding new trees that predict the residuals or errors of prior trees that are then combined with previous trees to make the final prediction.  It is called gradient boosting because it uses a gradient descent algorithm to minimize the loss when adding new models~\cite{10.1145/2939672.2939785}.

\subsubsection{LightGBM} \label{subsubsection:lightgbm}
The last ensemble method explained in this work is LightGBM \cite{10.5555/3294996.3295074}, a powerful gradient boosting framework that is highly scalable and efficient even for large data sets thanks to optimizations that enable faster training times and memory efficiency. 

LightGBM is an optimized implementation of GBDT. One of the main problems is that for data sets with large number of instances and features, GBDT needs to scan all the data instances to estimate the information gain of all possible split points. The solution, proposed in \cite{10.5555/3294996.3295074},  is two-fold,  namely, Gradient-based One-Sided Sampling (GOSS) and Exclusive Feature Bundling (EFB).
GOSS excludes a significant proportion of data instances with small gradients, and only uses the rest to estimate the information gain. EFB bundles mutually exclusive features to reduce the number of features.
With these two optimizations, the training process is sped up while maintaining almost the same accuracy.

\section{Results and discussion} \label{section:results}

This Section shows that ML models can achieve a high performance at predicting the value of the form factor for both cases, a single QM and two-coupled QMs.
In particular, it must be emphasized that, for the case of a single QM, almost perfect results can be obtained, with a $R^2$-score $\sim 0.998$ (see Table \ref{table:model_performance_single}). Thus, a trained lightweight machine learning model such as LightGBM (LGBM) can be used for faster calculations of form factor for this particular case instead of the slower numerical calculation.

The ML algorithms listed in previous sections were trained sequentially over the same training sets and evaluated over the testing sets. The data was split between training and test sets following a proportion of $2/3$ and $1/3$ and a fixed random state to be able to reproduce the experiment\footnote{The code is available at \url{https://github.com/carlos-hernani/QuantumMemristor}}. 

As mentioned in Section \ref{section:methodology}, different ML methods achieved high $R^2$-score values, most of them ensemble methods. 
The reason for choosing LightGBM over the other predictors shown at Tables \ref{table:model_performance_single} and \ref{table:model_performance_coupled} (see highlighted \textbf{LGBM} in bold) is that LightGBM has almost the highest value of our performance metric and faster training times. Additionally, LightGBM has other interesting features that make it our preferable option. First, as said before, the faster training allows us to test several iterations of the ML pipeline efficiently. Next, the high scalability thanks to the support of parallel, distributed and GPU learning with lower memory usage. This enhances the feasibility of training large data sets, which is expected for future research where three or more QMs might be coupled. While other models may keep reasonable scores, the training time will worsen greatly.
See Tables \ref{table:model_performance_single} and \ref{table:model_performance_coupled} for a comparison between models both in metric scores and time taken for training. 

\begin{figure}
    \centering
    \includegraphics[width=0.65\linewidth]{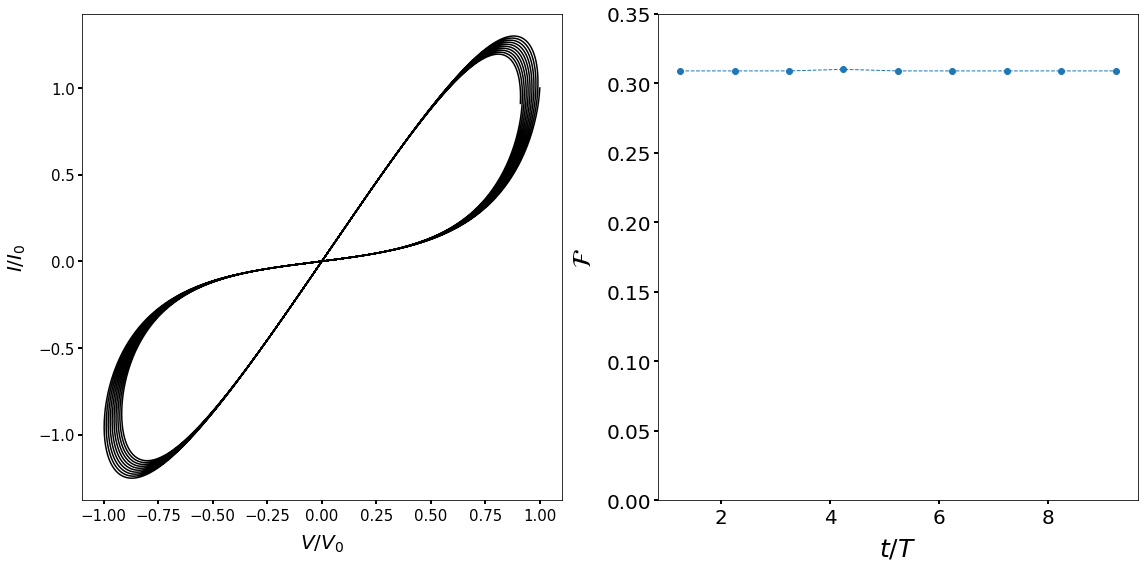}
    \caption{(a) Hysteresis curve of a single quantum memristor in the optimal configuration over 10 periods of the driving signal. (b) Form factor of the associated hysteresis curve, the form factor stays constant over time since it is invariant to decay. Time is in units of the period of the driving signal $T = 2\pi/\omega_{\ell}$}
    \label{fig.5}
\end{figure}

For the case of a single quantum memristor, we find that the optimal configuration that yields the highest value of form factor, corresponds to $\theta=\pi/2$, $\phi=1.5309$ and $\lambda=2.1387$. Fig.~\ref{fig.5} shows the corresponding hysteresis curve and form factor over time of the optimal case for a single memristor. This case is characterized by a very slow decay and stable hysteresis curve.

\begin{figure}
    \centering
    \includegraphics[width=0.7\linewidth]{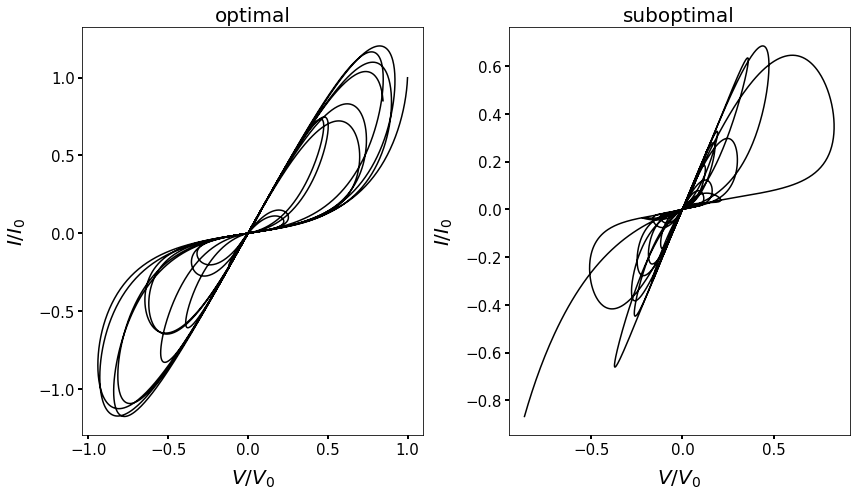}
    \caption{Pinched hysteresis curves of identical quantum memristors as a function of time for 10 periods of the driving signal. The optimal configuration, labeled as ‘optimal’, is compared with the configuration that yields the smallest form factor over time, denoted as ‘sub-optimal’. Time is in units of the period of the driving signal $T = 2\pi/\omega_{\ell}$}
    \label{fig.6}
\end{figure}

For the case of coupled quantum memristors the hysteresis loop in the I/V plane can change significantly over each period of the driving signal, and it is nontrivial to compare the dynamics corresponding to different configurations. 
We consider the optimal configuration, derived from the LightGBM model and which maximizes the form factor and compare it with the configuration that yields the lowest form factor according to our models, which we call sub-optimal configuration. The hysteresis curves of the optimal and sub-optimal cases are shown in Fig.~\ref{fig.6}. As can be seen, the optimal configuration preserves much better the shape of the memristive hysteresis curve when compared with the sub-optimal case. It is interesting that for the optimal configuration of the coupled case, the value of $lambda$ and $\phi$ coincides with the case of a single quantum memristor. 

\begin{table}[h]
 \caption{Machine learning model performance for a single QM.}
\label{table:model_performance_single}
\centering
    \begin{tabular}{cccccc} \toprule
    {} & {$Model$} & {Adjusted R-Squared} & {R-Squared} & {RMSE} & {Time Taken ($s$)}  \\ \midrule
    1 & GaussianProcess & 0.9998 & 0.9998 & 0.001028 & 0.103855 \\
    2 & NuSVR & 0.9992 & 0.9993 & 0.002289 & 0.435859 \\
    3 & HistGradientBoosting & 0.9980 & 0.9980 & 0.003729 & 0.379204 \\
    4 & \textbf{LGBM} & 0.9980 & 0.9980 & 0.003751 & \textbf{0.181819} \\ \midrule
    5 & ExtraTrees & 0.9975 & 0.9975 & 0.004212 & 0.127828 \\
    6 & RandomForest & 0.9970 & 0.9970 & 0.004577 & 0.198217 \\
    7 & Bagging & 0.9959 & 0.9959 & 0.005321 & 0.024523 \\
    8 & XGBoost & 0.9951 & 0.9951 & 0.005863 & 0.139264 \\
    9 & KNeighbors & 0.9942 & 0.9942 & 0.006346 & 0.004720 \\
    10 & GradientBoosting & 0.9933 & 0.9933 & 0.006830 & 0.112058 \\
    \bottomrule
\end{tabular}

\end{table}

\begin{figure}
    \centering
    \includegraphics[width=1\linewidth]{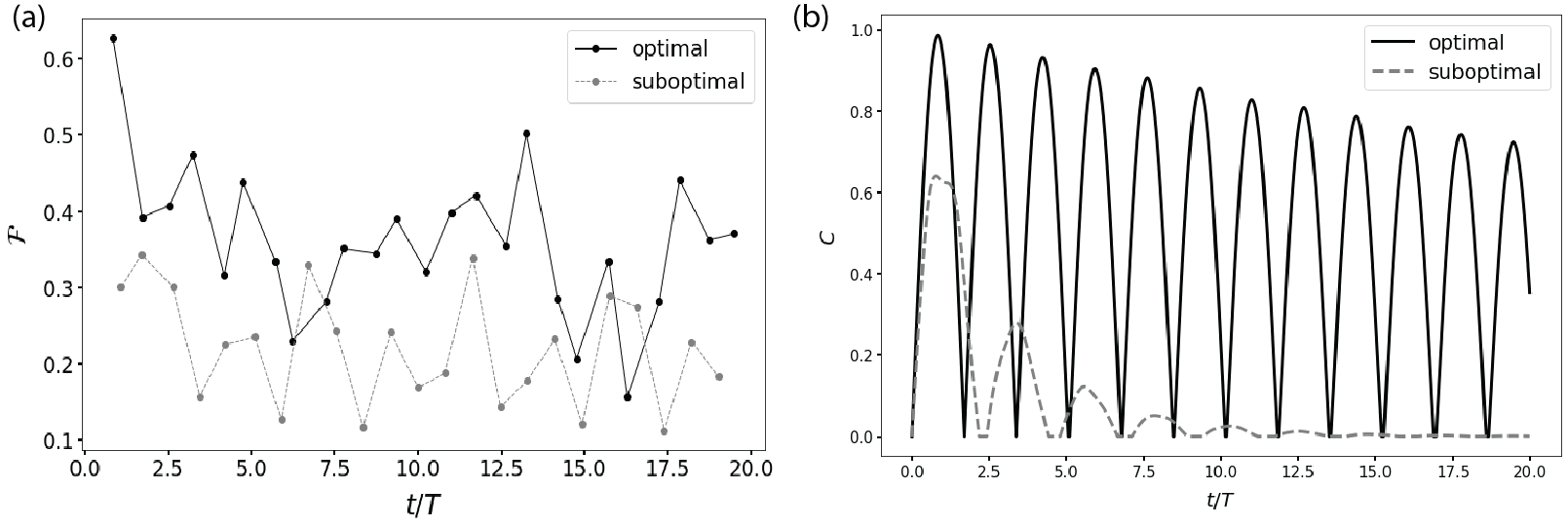}
    \caption{(a) Form factor of the hysteresis curves of Fig.~\ref{fig.6} as a function of time. (b) Entanglement between the quantum memristors measured by concurrence as a function of time for 20 periods of the driving signal $T = 2\pi/\omega_{\ell}$. The form factor may change significantly between periods, but the optimal configuration attains a higher form factor over the course of the evolution. Also, the optimal configuration, displays high and long lasting entanglement, conversely, the sub-optimal configuration  displays smaller values of concurrence which decay much faster.}
    \label{fig.7}
\end{figure}

In Fig.~\ref{fig.7}~(a) we show the comparison of the form factor calculated for both the optimal and sub-optimal configurations over the course of 20 periods of the driving signal. The results clearly demonstrate that the form factor for the optimal configuration consistently exceeds that of the sub-optimal configuration throughout the entire duration of the system’s evolution.

In addition, we examine the quantum correlations generated in the system by measuring the concurrence between the quantum memristors. Concurrence is an entanglement monotone,  defined for a mixed state of two two-level systems. It is calculated using the following formula:
\begin{equation}
C(\rho) = \max\big(0, \epsilon_1 - \epsilon_2 - \epsilon_3 - \epsilon_4 \big)
\end{equation}

where $\epsilon_i$ are the eigenvalues in decreasing order of the Hermitian matrix $\sqrt{\sqrt{\rho} \tilde{\rho} \sqrt{\rho}}$, with $\tilde{\rho} =\big( \sigma_y\otimes \sigma_y  \big) \rho^*  \big( \sigma_y\otimes \sigma_y  \big) $. Here, complex conjugation is taken in the eigenbasis of $\sigma_z$. Fig.~\ref{fig.7}~(b) shows the concurrence between the quantum memristors as a function of time for 20 periods of the driving signal, where we compare the optimal and sub-optimal configuration. Notice that  maximizing  the memory effects as measured by the form factor leads to high entanglement during the time evolution of the system. In contrast, the configuration that minimizes the form factor has significantly less entanglement which decays much faster than the optimal configuration. This is remarkable since it means that maximizing the form factor leads to a case with very high quantum correlations that can persist in high values for several periods of the driving signal. This is a separate and stronger result of the relation between memristive behavior and quantum correlations in coupled quantum memristors than that reported in \cite{PhysRevA.104.062605}.

\begin{table}[h]
\caption{Machine learning model performance for two-coupled QMs.}
\label{table:model_performance_coupled}   
\centering
    \begin{tabular}{cccccc} \toprule
    {} & {$Model$} & {Adjusted R-Squared} & {R-Squared} & {RMSE} & {Time Taken ($s$)}  \\ \midrule
    1 & ExtraTrees  & 0.97 & 0.97 & 0.01 & 1.13  \\
    2 & XGBoost  & 0.97  & 0.97 & 0.01  & 0.22  \\
    3 & RandomForest  & 0.97  & 0.97 & 0.01  & 1.74  \\
    4 & \textbf{LGBM}  & 0.97  & 0.97 & 0.01  & \textbf{0.12}  \\ \midrule
    5 & Bagging  & 0.96   & 0.96 & 0.01   & 0.19  \\
    6 & HistGradientBoosting  & 0.96  & 0.96 & 0.01  & 0.56  \\
    7 & DecisionTree  & 0.94  & 0.94 & 0.01  & 0.05  \\
    8 & ExtraTree  & 0.92  & 0.92 & 0.01  & 0.03  \\ 
    9 & KNeighbors  & 0.85  & 0.85 & 0.02  & 0.05  \\
    10 & GradientBoosting & 0.83   & 0.83 & 0.02   & 0.57  \\\bottomrule
\end{tabular}

\end{table}

\section{Conclusions}  \label{section:conclusions}

We have presented a ML approach to accurately model the memristivity for maximizing the memristive behavior of systems formed by a single QM and two coupled QMs. The high performance achieved by different ML models lays bare the robustness of the proposed approach, i.e., predicting and optimizing the memristive behavior is attainable and reliable. 

An additional advantage of ML stems from its capacity to allow for generalizations when exploring the parameter space. Therefore, its application surpasses the usual approach of analyzing and understanding the system that produces a given form factor since ML can generate the parameter configuration for a given form factor, means that is easy to check different hypothesis and the study of extremal case, for example maximizing memristive properties. This also involves a considrable reduction in computational costs.

Interestingly, in the optimal configuration, we note that the form factor surpasses $0.5$. This is the maximal value for symmetric hysteresis loops, i.e. both lobs with the same shape, revealing that for optimal configurations the system abandon a symmetric dynamics. We also observe that the amount of quantum correlation in the optimal case increases, yielding peaks close to 1. This suggests that, during the dynamics in the optimal case, the coupled memristor periodically migrates from unentangled to highly entangled states.

An extension to tripartite and multipartite quantum memristors, where different kinds of quantum correlations emerge, may have to be carefully considered. Moreover, due to the non-trivial enhancement of quantum correlations, it looks promising to study the influence of memristor-based protocols on quantum reservoir computing and quantum neural networks.

\section*{Acknowledgments}
This work was supported in part by the Valencian Government grant with reference number CIAICO/2021/184 and the Spanish Ministry of Economic Affairs and Digital Transformation through the QUANTUM ENIA project call – Quantum Spain project, and the European Union through the Recovery, Transformation and Resilience Plan – NextGenerationEU within the framework of the Digital Spain 2025 Agenda. We are grateful for the support of the Scientific Calculus' RES node of University of València for their service. Finally, F.A.-A. acknowledges the financial support from Agencia Nacional de Investigaci\'on y Desarrollo (ANID): Subvenci\'on a la Instalaci\'on en la Academia SA77210016, Fondecyt Regular 1231174, Financiamiento Basal para Centros Cient\'ificos y Tecnol\'ogicos de Excelencia AFB 220001.

\bibliographystyle{MSP}
\bibliography{references}

\begin{thebibliography}{10}
\providecommand{\url}[1]{\texttt{#1}}
\providecommand{\urlprefix}{URL }

\bibitem{1083337}
L.~Chua,
\newblock \emph{IEEE Transactions on Circuit Theory} \textbf{1971}, \emph{18},
  5 507.

\bibitem{1454361}
L.~Chua and S.~M. Kang,
\newblock \emph{Proceedings of the IEEE} \textbf{1976}, \emph{64}, 2 209.

\bibitem{doi:10.1143/JPSJ.12.570}
R.~Kubo,
\newblock \emph{Journal of the Physical Society of Japan} \textbf{1957},
  \emph{12}, 6 570.

\bibitem{Strukov2008}
D.~B. Strukov, G.~S. Snider, D.~R. Stewart et~al.,
\newblock \emph{Nature} \textbf{2008}, \emph{453}, 7191 80.

\bibitem{Barrios2019}
G.~A. Barrios, J.~C. Retamal, E.~Solano et~al.,
\newblock \emph{Scientific Reports} \textbf{2019}, \emph{9}, 1 12928.

\bibitem{schuman2017survey}
C.~D. Schuman, T.~E. Potok, R.~M. Patton et~al.,
\newblock A survey of neuromorphic computing and neural networks in hardware,
  \textbf{2017}.

\bibitem{Wang2017}
Z.~Wang, S.~Joshi, S.~E. Savel'ev et~al.,
\newblock \emph{Nature Materials} \textbf{2017}, \emph{16}, 1 101.

\bibitem{doi:10.1080/00018732.2010.544961}
Y.~V. Pershin and M.~D. Ventra,
\newblock \emph{Advances in Physics} \textbf{2011}, \emph{60}, 2 145.

\bibitem{Li_2018}
Y.~Li, Z.~Wang, R.~Midya et~al.,
\newblock \emph{Journal of Physics D: Applied Physics} \textbf{2018},
  \emph{51}, 50 503002.

\bibitem{Markovic2020}
D.~Markovi{\'{c}}, A.~Mizrahi, D.~Querlioz et~al.,
\newblock \emph{Nature Reviews Physics} \textbf{2020}, \emph{2}, 9 499.

\bibitem{Islam_2019}
R.~Islam, H.~Li, P.-Y. Chen et~al.,
\newblock \emph{Journal of Physics D: Applied Physics} \textbf{2019},
  \emph{52}, 11 113001.

\bibitem{GYONGYOSI201951}
L.~Gyongyosi and S.~Imre,
\newblock \emph{Computer Science Review} \textbf{2019}, \emph{31} 51.

\bibitem{Arute2019}
F.~Arute, K.~Arya, R.~Babbush et~al.,
\newblock \emph{Nature} \textbf{2019}, \emph{574}, 7779 505.

\bibitem{doi:10.1126/science.abe8770}
H.-S. Zhong, H.~Wang, Y.-H. Deng et~al.,
\newblock \emph{Science} \textbf{2020}, \emph{370}, 6523 1460.

\bibitem{PhysRevLett.127.180501}
Y.~Wu, W.-S. Bao, S.~Cao et~al.,
\newblock \emph{Phys. Rev. Lett.} \textbf{2021}, \emph{127} 180501.

\bibitem{doi:10.1126/sciadv.abi7894}
A.~Deshpande, A.~Mehta, T.~Vincent et~al.,
\newblock \emph{Science Advances} \textbf{2022}, \emph{8}, 1 eabi7894.

\bibitem{morvan2023phase}
A.~Morvan, B.~Villalonga, X.~Mi et~al.,
\newblock Phase transition in random circuit sampling, \textbf{2023}.

\bibitem{PhysRevA.104.062605}
S.~Kumar, F.~A. C\'ardenas-L\'opez, N.~N. Hegade et~al.,
\newblock \emph{Phys. Rev. A} \textbf{2021}, \emph{104} 062605.

\bibitem{PhysRevApplied.18.034004}
S.~Kumar, F.~C\'ardenas-L\'opez, N.~Hegade et~al.,
\newblock \emph{Phys. Rev. Appl.} \textbf{2022}, \emph{18} 034004.

\bibitem{10.1063/5.0020014}
D.~Marković and J.~Grollier,
\newblock \emph{Applied Physics Letters} \textbf{2020}, \emph{117}, 15, 150501.

\bibitem{Pfeiffer2016-oj}
P.~Pfeiffer, I.~L. Egusquiza, M.~Di~Ventra et~al.,
\newblock \emph{Scientific Reports} \textbf{2016}, \emph{6}, 1 29507.

\bibitem{10.1063/1.5036596}
M.~Sanz, L.~Lamata and E.~Solano,
\newblock \emph{APL Photonics} \textbf{2018}, \emph{3}, 8, 080801.

\bibitem{ma13040864}
T.~Gonzalez-Raya, J.~M. Lukens, L.~C. Céleri et~al.,
\newblock \emph{Materials} \textbf{2020}, \emph{13}, 4.

\bibitem{PhysRevApplied.6.014006}
S.~N. Shevchenko, Y.~V. Pershin and F.~Nori,
\newblock \emph{Phys. Rev. Appl.} \textbf{2016}, \emph{6} 014006.

\bibitem{PhysRevApplied.10.014013}
S.~N. Shevchenko and D.~S. Karpov,
\newblock \emph{Phys. Rev. Appl.} \textbf{2018}, \emph{10} 014013.

\bibitem{PhysRevApplied.2.034011}
S.~Peotta and M.~Di~Ventra,
\newblock \emph{Phys. Rev. Appl.} \textbf{2014}, \emph{2} 034011.

\bibitem{Salmilehto2017-zi}
J.~Salmilehto, F.~Deppe, M.~Di~Ventra et~al.,
\newblock \emph{Scientific Reports} \textbf{2017}, \emph{7}, 1 42044.

\bibitem{e25081134}
S.~Stremoukhov, P.~Forsh, K.~Khabarova et~al.,
\newblock \emph{Entropy} \textbf{2023}, \emph{25}, 8.

\bibitem{Tang2023}
J.-L. Tang, G.~Alvarado~Barrios, E.~Solano et~al.,
\newblock \emph{Entropy} \textbf{2023}, \emph{25}, 5.

\bibitem{PhysRevApplied.17.024056}
A.~Norambuena, F.~Torres, M.~Di~Ventra et~al.,
\newblock \emph{Phys. Rev. Appl.} \textbf{2022}, \emph{17} 024056.

\bibitem{Spagnolo2022-ie}
M.~Spagnolo, J.~Morris, S.~Piacentini et~al.,
\newblock \emph{Nature Photonics} \textbf{2022}, \emph{16}, 4 318.

\bibitem{GAO2022100007}
J.~Gao, X.-W. Wang, W.-H. Zhou et~al.,
\newblock \emph{Chip} \textbf{2022}, \emph{1}, 2 100007.

\bibitem{PhysRevA.61.052306}
V.~Coffman, J.~Kundu and W.~K. Wootters,
\newblock \emph{Phys. Rev. A} \textbf{2000}, \emph{61} 052306.

\bibitem{PhysRevLett.113.100503}
Y.-K. Bai, Y.-F. Xu and Z.~D. Wang,
\newblock \emph{Phys. Rev. Lett.} \textbf{2014}, \emph{113} 100503.

\bibitem{PhysRevApplied.8.024030}
K.~Fujii and K.~Nakajima,
\newblock \emph{Phys. Rev. Appl.} \textbf{2017}, \emph{8} 024030.

\bibitem{FERRERSANCHEZ2022105967}
A.~Ferrer-Sánchez, J.~Bagan, J.~Vila-Francés et~al.,
\newblock \emph{Oral Oncology} \textbf{2022}, \emph{132} 105967.

\bibitem{liu2023summary}
Y.~Liu, T.~Han, S.~Ma et~al.,
\newblock Summary of chatgpt/gpt-4 research and perspective towards the future
  of large language models, \textbf{2023}.

\bibitem{rombach2022highresolution}
R.~Rombach, A.~Blattmann, D.~Lorenz et~al.,
\newblock High-resolution image synthesis with latent diffusion models,
  \textbf{2022}.

\bibitem{PhysRevResearch.3.013167}
A.~Martin, B.~Candelas, A.~Rodr\'{\i}guez-Rozas et~al.,
\newblock \emph{Phys. Rev. Res.} \textbf{2021}, \emph{3} 013167.

\bibitem{e25020323}
Y.~Ding, J.~Gonzalez-Conde, L.~Lamata et~al.,
\newblock \emph{Entropy} \textbf{2023}, \emph{25}, 2.

\bibitem{Wang2021-vr}
R.~Wang, C.~Hernani-Morales, J.~D. Mart{\'\i}n-Guerrero et~al.,
\newblock \emph{Quantum Science and Technology} \textbf{2021}, \emph{7}, 1
  015010.

\bibitem{Flores}
C.~Flores-Garrigós, J.~Vicent-Camisón, J.~Garces et~al.,
\newblock \emph{Applied Sciences} \textbf{2021}, \emph{11} 11754.

\bibitem{ferrersanchez2023gradientannihilated}
A.~Ferrer-Sánchez, J.~D. Martín-Guerrero, R.~R. de~Austri et~al.,
\newblock Gradient-annihilated pinns for solving riemann problems: Application
  to relativistic hydrodynamics, \textbf{2023}.

\bibitem{JOSEPHSON1962251}
B.~Josephson,
\newblock \emph{Physics Letters} \textbf{1962}, \emph{1}, 7 251.

\bibitem{PhysRevB.86.184514}
G.~Catelani, S.~E. Nigg, S.~M. Girvin et~al.,
\newblock \emph{Phys. Rev. B} \textbf{2012}, \emph{86} 184514.

\bibitem{IEEE2012Biolek}
Z.~Biolek, D.~Biolek and V.~Biolkova,
\newblock \emph{IEEE Trans. Circuits Syst., II, Exp. Briefs} \textbf{2012},
  \emph{59}, 9 607.

\bibitem{EL2014Biolek}
D.~Biolek, Z.~Biolek and V.~Biolkov{\'a},
\newblock \emph{Electronics Letters} \textbf{2014}, \emph{50}, 2 74.

\bibitem{NIPS1996_d3890178}
H.~Drucker, C.~J.~C. Burges, L.~Kaufman et~al.,
\newblock In M.~Mozer, M.~Jordan and T.~Petsche, editors, \emph{Advances in
  Neural Information Processing Systems}, volume~9. MIT Press, \textbf{1996} .

\bibitem{MOHAMMED2023757}
A.~Mohammed and R.~Kora,
\newblock \emph{Journal of King Saud University - Computer and Information
  Sciences} \textbf{2023}, \emph{35}, 2 757.

\bibitem{Ganaie_2022}
M.~Ganaie, M.~Hu, A.~Malik et~al.,
\newblock \emph{Engineering Applications of Artificial Intelligence}
  \textbf{2022}, \emph{115} 105151.

\bibitem{Dong2020}
X.~Dong, Z.~Yu, W.~Cao et~al.,
\newblock \emph{Frontiers of Computer Science} \textbf{2020}, \emph{14}, 2 241.

\bibitem{decision_trees}
L.~Rokach and O.~Maimon,
\newblock \emph{Decision Trees}, volume~6, 165--192,
\newblock Springer, \textbf{2005}.

\bibitem{Breiman2001}
L.~Breiman,
\newblock \emph{Machine Learning} \textbf{2001}, \emph{45}, 1 5.

\bibitem{Geurts2006}
P.~Geurts, D.~Ernst and L.~Wehenkel,
\newblock \emph{Machine Learning} \textbf{2006}, \emph{63}, 1 3.

\bibitem{10.1145/2939672.2939785}
T.~Chen and C.~Guestrin,
\newblock In \emph{Proceedings of the 22nd ACM SIGKDD International Conference
  on Knowledge Discovery and Data Mining}, KDD '16. Association for Computing
  Machinery, New York, NY, USA,
\newblock ISBN 9781450342322, \textbf{2016} 785–794.

\bibitem{10.5555/3294996.3295074}
G.~Ke, Q.~Meng, T.~Finley et~al.,
\newblock In \emph{Proceedings of the 31st International Conference on Neural
  Information Processing Systems}, NIPS'17. Curran Associates Inc., Red Hook,
  NY, USA,
\newblock ISBN 9781510860964, \textbf{2017} 3149–3157.

\end{thebibliography}

\end{document}